\documentclass[letterpaper,11pt]{article}
\usepackage{tabularx} 
\usepackage{amsmath}  
\usepackage{graphicx} 
\usepackage[margin=1in,letterpaper]{geometry} 
\usepackage{revsymb}

\usepackage[final]{hyperref} 
\hypersetup{
	colorlinks=true,       
	linkcolor=blue,        
	citecolor=blue,        
	filecolor=magenta,     
	urlcolor=blue         
}
\usepackage{blindtext}

\usepackage{graphicx}  
\usepackage{subfigure}
\usepackage{multirow}

\usepackage{fancyhdr}
\usepackage{longtable}
\usepackage{parskip}
\usepackage[T1]{fontenc}
\usepackage{dcolumn}   

\usepackage{bm}        
\usepackage{amsfonts}  
\usepackage{amsmath}   
\usepackage{amssymb}   

\usepackage{mathrsfs}  
\usepackage{amsmath,amsfonts}
\usepackage{tcolorbox}
\usepackage{titlesec}

\usepackage{url}

\usepackage{graphicx}  
\usepackage{subfigure}
\usepackage{multirow}
\usepackage{authblk}

\usepackage{fancyhdr}
\usepackage{longtable}
\usepackage{parskip}
\usepackage[T1]{fontenc}
\usepackage{dcolumn}   
\usepackage{braket}

\usepackage{bbm}
\usepackage{tikz}
\usepackage{caption}
\usepackage{hyperref}

\usepackage{collectbox}

\usepackage{titlesec}

\usepackage{url}

\date{} 

\begin{document}

\title{An examination of the hierarchy problem \\beyond the Standard Model}
\author{Seyed Khaki\footnote{Academy of Fundamental Studies, Munich 80637, Germany\\ Email: \href{mailto:s.khaki@afundas.com}{s.khaki@afundas.com}}}

\maketitle
\setlength{\parindent}{3ex}
\begin{abstract}
As the Higgs field is a weak isospin doublet of the SU(2) symmetry, the Standard Model requires any symmetry solution to the Higgs hierarchy problem to be SU(2) invariant, a constraint on the type of the symmetry. However, the hierarchy problem is about the size. The size of SU(2) for the Higgs boson can be calculated by $|$SU$_2(\ell)|=\ell^3-\ell$, having the Higgs mass $M_H=1/\ell_{H}\approx125$ GeV. To find the origin of the relative smallness of the Higgs mass in Planck units, alternatively, we search for the origin of such a large order assuming that it stems from an unknown field theory X beyond the Standard Model. Accordingly, this order, which corresponds to the quantum of the Higgs field, should determine the order of quantum/core of X symmetry, its automorphism group. We calculate $|$Aut(X)$|\approx8.2\times 10^{53}$, close to the order of the Monster sporadic group, $|\mathbb M|\approx 8.1\times 10^{53}$, the automorphism group of the Monster CFT, which we therefore conjecture to be X. To examine this conjecture, we calculate the mass of a scalar boson whose SU(2) order is determined by $|\mathbb M|$, observing a 125.4 GeV boson mass and a 245.7 GeV VEV. The Monster CFT does not have any spin-1 operators and Kac-Moody symmetry. Therefore, based on the CFT/(A)dS correspondences, it only describes pure gravity without the gauge fields. In search of a gauge theory candidate, we promote SU(2) (double cover of SO(3)), to SO($d$), and show that the same $\mathbb M$-symmetric vacuum configuration reaches the Planck mass of quantum gravity precisely at $d=32$ (with 99\% accuracy). Then, the spin-1 boson mass of the eligible gauge candidates, SO(32) and $E_8\times E_8$, is calculated to be 80.9 GeV. Further, several pieces of evidence are provided supporting the conjecture.
\end{abstract}

\section{Introduction}
Despite several smart proposals for the hierarchy problem (see \cite{koren2020hierarchy} for a review), there is no agreement yet on the origin of the large hierarchy between the electroweak scale of the Higgs mass and the Planck mass, $\frac{M_P}{M_H}\approx 1.95\times 10^{16}$. In the Standard Model, the scalar Higgs field is a weak isospin doublet of the SU(2) symmetry. On the one hand, the Standard Model, including its SU(2) symmetry, is incredibly successful in predicting many aspects of elementary particles and their interactions. On the other hand, SU(2) alone could not explain the origin of the Higgs mass's smallness relative to the Planck mass. Thus, a solution to the hierarchy problem must be necessarily SU(2)-invariant to accord with the Standard Model. This puts constraints on the type of symmetry. However, the hierarchy problem is a question about the size. Fortunately, due to the hard work of many people worldwide, the Higgs mass is measured and we now know the size of the quantum of the Higgs field. Hence, we can estimate the size of SU(2) elements/transformations for the Higgs boson but do not know its source. We guess it probably originates from beyond the Standard Model since we are clueless inside it. Thus, our approach assumes the existence of an unknown complementary symmetry that, first, conforms to and coexists with SU(2), second, it restricts/determines the size of SU(2) for the quantum of the field. Particularly, we assume that the quantum/core symmetry (automorphism group) of an unknown field theory X restricts the size of the SU(2) symmetry of the quantum of the Higgs field, and therefore, limits the size of the Higgs boson. In the next section \ref{section:motivation}, having the Higgs mass as the input, we observe that such an assumption yields $|$Aut(X)$|\approx 8.2\times 10^{53}$. This is considerably close to the order of the Monster group ($|\mathbb M|\approx 8.1 \times 10^{53}$), the largest sporadic simple group, which is the automorphism group of the Monster conformal field theory (CFT). Consequently, we conjecture that X is the Monster CFT. To examine this conjecture, we calculate the mass of a scalar field boson with SU(2) symmetry, whose order is determined by $|\mathbb M|$. We observe a 125.4 GeV mass for the boson and 245.7 GeV for its corresponding VEV. We present several pieces of evidence explaining why the Monster CFT is physically eligible to be the underlying theory and why its appearance is neither unexpected nor a coincidence. 

The Monster CFT does not have any spin-1 operators and Kac-Moody symmetry. Therefore, based on the CFT/(A)dS correspondences \cite{strominger2001ds} \cite{maldacena1999large}, it only describes pure gravity which does not include the gauge fields. To find a gauge theory candidate that describes the gauge fields, in the see cond part \ref{section:gauage}, we promote the 3-dimensional SU(2) (double cover of SO(3)), to $d$-dimensional SO($d$), and show that the same $\mathbb M$ vacuum configuration reaches the Planck mass of quantum gravity precisely at $d=32$ (with 99\% accuracy). Then, the spin-1 boson mass of the eligible gauge candidates, SO(32) and $E_8\times E_8$, is calculated to be 80.9 GeV. We emphasize that this work does not aim to prove the correctness of its conjecture but rather to report evidence motivating future comprehensive studies to investigate its validity and potential.
\section{Motivation}
\label{section:motivation}
In the Standard Model, the scalar Higgs field is a weak isospin doublet of the SU(2) symmetry. SU(2) is a 3-dimensional Lie group possessing 3 generators, equal to that of the 3-dimensional SO(3) group (SU(2) is indeed the universal covering group of SO(3)). This can also be seen in the order of the SU(2) group over finite fields, where in $|SU_2(\ell)|= \ell^3-\ell$ \footnote{In the mathematics literature, the common notation is $|$SU$_2(q)|= q^3-q$, where the integer $q$ is the size of the underlying complex scalar field $\mathbb F_q$.} the leading term is a cube of $\ell$, the characteristic length of the underlying scalar field. The characteristic length of the Higgs field is the Compton wavelength of the Higgs boson, $\ell_H\approx 1.943\times 10^{16}\: \ell_P$ ($\frac{\ell_H}{\ell_P}= \frac{M_P}{M_H}\approx 1.943\times 10^{16}$ \footnote{The reduced Planck mass ($\approx 2.435\times 10^{18}$) is employed throughout this paper.}). This means that it consists of approximately $1.95\times 10^{16}$ Planck elements/length (i.e., its value in the Planck units). Consequently, the total number of 3-dimensional SU(2) transformations between these Planck elements is approximately its cube, $(1.943\times 10^{16})^3\approx 7.3\times 10^{48}$ \footnote{This order can also be interpreted as an estimation for the number of Planck cubes (cubes of length $\ell_P$) that can be embedded in a Higgs cube (cubes of length $\ell_H$).}. Since we do not know the origin of this order, let us assume that it originates from an unknown theory X beyond the Standard Model that limits $|$SU(2)$|$ and therefore determines a specific value for $\ell_H$. Because this order stems from the size of the quantum of the Higgs field, based on our assumption, it estimates the size of the quantum/core symmetry of X, its automorphism group. In other words, we use the experimentally known $\ell_H$ to calculate the restricted size of $|$SU(2)$|$ for the Higgs boson, which based on our assumption is dictated by $|$Aut(X)$|$. Thus, the naive estimation is $|$Aut(X)$|\approx 7.4\times 10^{48}$.

The mentioned 3 dimensions/generators correspond to the electroweak scale of the Higgs mass with SU(2) symmetry. At higher energies, the symmetry is larger, and the number of dimensions and DOF are higher. In our naive estimation, which calculates the ratio of the final state of the Higgs scale to the initial state of the Planck scale, the Planck scale arises at an infinitely large number of dimensions, resulting in $\frac{\ell_H}{\ell_P}\approx \frac{|Aut(X)|^\frac{1}{3}}{|Aut(X)|^\frac{1}{\infty}}=|Aut(X)|^\frac{1}{3}$, hence, $|Aut(X)|\approx (\frac{\ell_H}{\ell_P})^3$. However, for the Planck scale, where the effects of gravity are significant, string theory introduces the SO(32) group as a candidate gauge group of quantum gravity (in type I and one of heterotic superstring theories). This suggests that the total number of transformations among $\ell$ elements (field of length $\ell$) is about $\ell^{32}$. Assuming that 32 is the right dimension for the Planck scale, $\ell_P^{\:32}$ becomes an estimation for the total number of transformations over the field of Planck length, and hence for the order of Aut(X). Accordingly, the naive estimation improves to
\begin{equation}
\label{eq:1}
    \frac{|Aut(X)|^\frac{1}{3}}{|Aut(X)|^\frac{1}{32}} \approx \frac{\ell_H}{\ell_P}
\end{equation}
The numerators and denominators are the size of the scalar field at two different eras. The Planck epoch, where the field length is $|$Aut(X)$|^\frac{1}{32}$, and the (post) electroweak era, where the field length is $|$Aut(X)$|^\frac{1}{3}$. The 32 and 3 powers are borrowed from string theory and the Standard Model, respectively. From Eq. \ref{eq:1},
\begin{equation}
\label{eq:2}
    |Aut(X)|\approx (\frac{\ell_H}{\ell_P})^{3.31}\approx 8.2\times 10^{53}
\end{equation}
Strikingly, this is very close to the order of the largest sporadic simple group, the Monster group $|\mathbb M|\approx 8.1 \times 10^{53}$. Remarkably, in 1984, a conformal field theory whose automorphism group is $\mathbb M$ is constructed \cite{frenkel1984natural}. Apparently, this is the only known field theory with $\mathbb M$ symmetry. Therefore, we conjecture that X is the Monster CFT.

The Monster CFT is an orbifold theory. Before the employment of orbifolds in string theory, mathematicians studied asymmetric orbifold constructions to probe the properties of the largest sporadic group, the Monster group \cite{conway1979monstrous}. Construction of the Monster CFT, which emerged independently in the context of string compactification, can be regarded as the earliest example of an orbifold CFT. Since then, many studies uncovered various properties of this theory by addressing different aspects of it (for developments see \cite{gannon2007moonshine}, for a recent review see \cite{harrison2022snowmass}), and investigated its physical implications in the context of the string theory. In the construction of the Monster CFT, the 26-$d$ target space of the bosonic string is compactified to 2-$d$. The space of 24 transverse dimensions is discretized by the Leech lattice and the CFT is obtained by gauging/orbifolding the $\mathbb{Z}_2$ symmetry. The resulting partition function is the modular function of weight zero $J(\tau)= j(\tau)-744$, where $j(\tau)= e^{-2\pi i \tau}+744+196884e^{2\pi i \tau}+21493760e^{4\pi i \tau}+...$ is the renowned modular invariant $j$-function.

\section{Results}
\label{section:Results}
\subsection{Scalar boson}
\label{section:scalar}
Motivated by the previous observation, let us assume that the SU(2) symmetry of a scalar field of length $\ell$ is restricted by the automorphism group of the Monster CFT, the Monster group. 
\begin{equation}
\label{eq:3}
   |SU_2(\ell)| \leq |\mathbb M| 
\end{equation}
At the upper bound
\begin{equation}
\label{eq:4}
   |SU_2(\ell)| \approx |\mathbb M| 
\end{equation}
Note that the LHS stems from the Standard Model and RHS arises from its beyond. Accordingly,
\begin{equation}
\label{eq:5}
   \ell^3 -\ell \approx 8.08 \times 10^{53}  
\end{equation}
Solving for $\ell$ provides the length $|\ell| \approx 9.314\times 10^{17}$. These $\ell$ elements propagate in 24 transverse (small compactified) dimensions because the central charge of the Monster CFT is 24. The energy of each element oscillating in 24 dimensions is equal to the energy of 24 elements oscillating in 1 dimension. That is, $\ell$ elements propagating in 1 dimension are equivalent (in terms of energy) to $\ell/24$ elements propagating in 24 dimensions. Thus, the equivalent field length (the Compton wavelength of its boson) considering the central charge is $\ell/24$, and the boson mass is its inverse, $24/\ell$ \footnote{This is the established impact of the central charge that appears in the numerator of the energy. For instance, to obtain the ground state energy of a CFT with $c = 24$, one calculates the sum of ground state energies of an infinite number of harmonic oscillators, $(1+2+3+...)/2 = \zeta(-1)/2 = -1/24$, and since they propagate in 24 transverse dimensions, multiply it by 24 yielding the known $L_0 = -1$. In general, for $c=24k$, $L_0=-c/24=-k$.}. Notably, this mass (which is obtained from the Compton wavelength of the particle) is equivalent to the mass of an open string in string theory as for an open string we have $m_s=1/\ell_s$. However, string theory suggests that such a boson should be a closed string that is made of two open strings attached, where for a closed string we have $m_s=2/\ell_s$. Hence, the boson mass becomes
\begin{equation}
\label{eq:6}
    m_{\ell}=\frac{2}{\ell/24}=\frac{48}{\ell}\approx 5.15\times10^{-17}
\end{equation}
Recovering the dimension by the reduced Planck mass $M_P\approx 2.435 \times10^{18} \rm\:GeV $, yields
\begin{equation}
\label{eq:7}
    M_{\ell}\approx 5.15\times10^{-17}\times M_P \approx 125.4 \rm\:GeV
\end{equation}
The essential observation is that this value is almost identical to the measured Higgs boson mass of $125.35$ GeV by CMS \cite{sirunyan2020measurement} and $125.22$ GeV by ATLAS \cite{aad2023measurement} in the Large Hadron Collider. 

\par \textbf{Interpretation} \\
Here, we offer several pieces of evidence supporting the conjecture, explaining why the Monster CFT is physically eligible to be the underlying theory and why its appearance is neither surprising nor a coincidence.
\begin{enumerate}
    \item Witten's contributions provide the first lines of motivation. In 2001, Witten argued if the quantum Hilbert space in de Sitter space has a finite dimension $N$, this gives a strong hint that Einstein’s theory with Lagrangian
\begin{equation}
\label{eq:8}
    I=-\frac{1}{8\pi G} \int  d^n x \sqrt{g} R - \Lambda \int d^n x \sqrt{g}
\end{equation}
    cannot be quantized and must be derived from a more fundamental theory that determines the possible values of $G \Lambda^{(n-2)/n}$ and $N$ as a nontrivial function of it \cite{witten2001quantum}. He reasons that the absence of a classical de Sitter limit suggests that the possible values of $N$ in string/M-theory are 'sporadic', rather than arising from 'infinite families'. Later in 2007, he proposed that the Monster CFT is very likely the dual CFT of 3D pure gravity in Anti-de Sitter space \cite{witten2007three}.
    \item The finite simple groups are the building blocks of symmetry. Therefore, it is natural to expect the symmetry of the quantum of a field (Higgs boson) to be a quantum of symmetry, one of the finite simple groups ($\mathbb M$).
    \item Due to the hard work of hundreds of mathematicians, the finite simple groups are classified into a couple of categories: infinite and finite \cite{wilson1999atlas}. The finite category contains the sporadic groups and the infinite category contains 3 classes, the cyclic groups $\mathbb Z_q$, the alternating groups $A_n$, and the groups of Lie type (including the classical Chevalley groups, e.g. SL$_{n+1}$($q$), the twisted Chevalley, e.g. SU$_{n+1}$($q$), the exceptional Chevalley groups, e.g. $E_8(q)$, and the Suzuki–Ree groups $^2G_2(q)$). The adjective 'infinite' refers to the fact that there is an infinite number of these groups in each class. For instance, selecting an arbitrary integer $n$ for SL$_n(q$), say $n=2$, we get an infinite number of groups SL$_2(2$), SL$_2(3$), SL$_2(4$), ..., with the order spectrum of [$6, \:24, \:120, ...$). Since the order spectra of the groups in the infinite category are unbounded above, there is no maximum order in the family to be introduced as the X symmetry, to be inserted in the RHS of Eq. \ref{eq:3}, and to restrict the scalar field's size. In other words, no matter which group one selects, there is always a larger group with a larger length and lower energy, leading to the absence of a natural cutoff. On the other hand, the order spectrum of the 26 sporadic groups in the finite category, [ $|M_{11}|=7920,\: |M_{12}|=95040, ..., \: |\mathbb M|\approx 8\times 10^{53}$], is bounded and has a natural maximum/cutoff $|\mathbb M|$, which introduces a natural length.
    \item It is well-established that the current value of the Higgs mass is critical \cite{buttazzo2013investigating}. Accordingly, it is natural to expect that the theory that explains Higgs mass should enjoy this feature of criticality. Remarkably, the Monster group is the largest sporadic group, and therefore, is the critical symmetry in the sporadic family. This criticality is evident in Eq.\eqref{eq:3}, where the Higgs mass is calculated from its upper bound. 
    \item The lower bound of Eq.\eqref{eq:3} is at the Planck mass, where $m_{\ell}=1$, which corresponds to $\ell=48$ (according to Eq. \ref{eq:6}), and hence, $|$SU$_2$(48)$|=48^3-48=110544$. This order is close to the other end of the sporadic order spectrum, the Mathieu groups order \footnote{The Monstrous moonshine development is recently extended by the link between the largest Mathieu group $M_{24}$ representation and the elliptic genus of $K3$ surfaces \cite{eguchi2011notes}. Although there are many aspects of the Mathieu moonshine that are mysterious and yet unclear, it ultimately refers to the massive sector too \cite{taormina2015twist}.}. Additionally, the smallest No. elements introduced above is $\ell/24$, which becomes 2 for $\ell=48$. Notably, based on the standard definition of the 'field' in mathematics, 2 is the smallest number of elements of a field. However, the field with a single element (first proposed in \cite{tits1956analogues}), is a well-known mathematical object that behaves like a finite field with one element, if such a field could exist. The collapse of the field concept below 2 elements nicely matches the current picture of physics, where it is believed that the typical concept of spacetime collapses below the Planck length, the length of the fundamental string (for more details and the original role of the field with one element see \cite{khaki2024original}).
    \item The next indication is evident in the Monster CFT's partition function (and also the partition function of CFTs with symmetry of other sporadic groups). In $J(\tau)=j(\tau)-744=e^{-2\pi i \tau}+196884e^{2\pi i \tau}+21493760e^{4\pi i \tau}+...$, there is no massless DOF, the theory ultimately refers to the massive sector, and consequently to the Higgs field, the origin of the mass. Further, according to the (A)dS/CFT correspondence \cite{strominger2001ds} \cite{maldacena1999large}, it describes a bulk theory of pure gravity whose source is merely the 'mass', revealing its relation with the Higgs field. 
    \item The Monster CFT satisfies the abovementioned requirement that its symmetry conforms to and
    coexists with SU(2). The Monster CFT is a modular invariant theory as its partition function is invariant under the action of the modular group SL$_2$($\mathbb Z$). Notably, SL$_2$($\mathbb Z$) has 3 generators similar to the SU(2) symmetry of the Higgs field. Note that SL$_2$($\mathbb Z$) is defined over the infinite field of integers $\mathbb Z$. Over the finite field of length $\ell$ (with $\mathbb Z_\ell$ symmetry), their orders are the same, SL$_2$($\ell$)$=\ell^3-\ell$. This order is also encoded in the central extension term $m^3-m$ of the Virasoro algebra, $[L_m, L_n]=(m-n)L_{m+n}+\frac{c}{12}(m^3-m)\delta_{m+n,0}$. In particular, over a field $k$ of characteristic $\ell$, the Witt algebra is defined to be the Lie algebra of derivations of the ring $k[z]/z^\ell$, where the Witt algebra is spanned by $L_m$ for $-1\leq m\leq \ell-2$.
    \item The Leech lattice, $\Lambda_{24}$, by which the Monster CFT is constructed, is so exceptional. Particularly, in dimension 24, it is the only lattice with no roots, universally optimal, and has the densest sphere packing \cite{cohn2017sphere}, implying that between all point configurations of the same density, it provides the minimum possible Gaussian energy \cite{cohn2022universal}. Thus, it is not surprising that it can be a preferred choice of nature due to its optimality.
    
    \item The phase transition of the electroweak spontaneous symmetry-breaking is a prediction of the proposed framework. In particular, according to the celebrated Lee-Yang theory \cite{yang1952statistical} \cite{lee1952statistical}, in the infinite-size limit of a finite-size system, when the complex zeros of the partition function become numerous and dense along a certain arc, a phase transition happens. In our system, such an arc is the circle of an underlying field $\phi=\phi_{0}$ exp($\frac{2\pi i n}{\ell}$), where $0\leq n\leq \ell-1$. The complex zeros are the discussed Planck elements (elements of Planck size) which become (con)dense when they accumulate to $|\mathbb M|^{1/3}$ amount, triggering the condensation.

    \item The Pariah family of the sporadic groups are of phenomenological interest for the masses that we are unaware of their origin, such as dark matter and neutrino masses. This is not specifically a reason but rather an implication that potentially fills an unknown theoretical gap. Particularly, the sporadic groups are divided into 2 separate families, namely the Monster family who are subquotients of the Monster group, and the Pariah family (consisting of the Lyons, O'Nan, Rudvalis, and 3 Janko groups) who are not subquotients of the Monster group (i.e., they have factors that do not divide the Monster order). Based on the observations, if the Monster (family) governs the Higgs field which is responsible for giving mass to the ordinary elementary particles, then one can naturally expect that each group within the Pariah family \footnote{as well as the Tits group which is sometimes regarded as the $27^{th}$ sporadic group because it is the only simple group that is a derivative of a group of Lie type and not precisely a group of Lie type} offer candidates for the mass generation of distinct types of particles as they are not involved in the Monster group. On the other hand, our current understanding of particle physics and cosmology suggests that there should be a distinct source of mass (except the Higgs boson) for the dark matter and the Neutrinos. Thus, this gap in our understanding can potentially be filled by the symmetries within the Pariah family.
    
    \item The existence of a superconformal field theory that sits exactly one dimension below the Monster CFT makes the Higgs mechanism conceivable. It is known that developing a non-zero VEV via the Higgs mechanism/tachyon condensation reduces the total rank by 1 unit. For instance, consider the SO($n$) theory whose field acquires a non-zero VEV. There exists a gauge transformation by which one can fix the components into the vector representation of the form ($cf$. \cite{suyama2002properties} Eq. 7.1)
\begin{equation}
\label{eq:9}
    \vec{\phi} = \frac{1}{C}\:\left(\begin{array}{c} VEV \\ 0 \\ \vdots \\ 0 \end{array}\right)_{n\:\times \:1}
\end{equation}
where $C$ is a constant. It can be observed that SO($n$) breaks to SO($n-1$). In string theory, this is an example of tachyon condensation, where, e.g., SO($n$) decays to SO($n-1$) (compactified on $S^1$) allowing the system to reduce its energy.

At the central charge 24 of the Monster CFT holomorphic factorization is possible and the total central charge is the sum of the central charges of the holomorphic CFT and its anti-holomorphic counterpart, i.e., $24+24=48$. Therefore, developing a nonzero VEV reduces it to 47, which is equivalent to a central charge of 23.5 for each (Anti)holomorphic sector. Remarkably, at the central charge 23.5, a superconformal field theory exists whose automorphism group is the double cover of the Baby Monster group, $\mathbb B$ \cite{hohn2010group}. Its vertex operator superalgebra is called the Shorter Moonshine module because its construction is based on the 23-dimensional 'Shorter Leech' lattice, $O_{23}$, playing a similar role as $\Lambda_{24}$ in the Monster CFT. The symmetry group of $\Lambda_{24}$ and $O_{23}$ are, respectively, the double covers of the first two Conway sporadic groups $Co_1$ and $Co_2$. Moreover, $\mathbb B$ and $Co_2$ are the subquestions and largest maximal subgroups of $\mathbb M$ and $Co_1$, respectively (this can be easily observed in a diagram of sporadic groups displaying subquestion relations). Notably, $\Lambda_{24}$ and $O_{23}$ are unimodular lattices without roots. A simple interpretation of unimodularity is that the volume of each lattice cell must be one, which is a normal well-understood property. However, the property of having no roots makes these lattices extremely exceptional such that under the critical dimension 26, $\Lambda_{24}$, $O_{23}$, and the odd Leech lattice $O_{24}$ are the only unimodular lattices without root \cite{conway2013sphere} \cite{conway1988low}.

Furthermore, in the construction of the Monster CFT, the 26-$d$ target space of the bosonic string is compactified to 2-$d$, where the space of 24 transverse dimensions is discretized by $\Lambda_{24}$. That is $26-24=2$. According to the Holographic principle, and particularly the (A)dS/CFT correspondences \cite{strominger2001ds} \cite{maldacena1999large}, this 2-$d$ CFT describes a 3-$d$ bulk spacetime, one dimension higher. On the other hand, if the transverse dimensions become 23-dimensional (e.g., discretized by $O_{23}$), the 26-$d$ target space of the bosonic string would be compactified to 3-$d$. Then, according to (A)dS/CFT correspondences, it would describe a 4-$d$ bulk spacetime.

\end{enumerate}

\subsection{VEV}
\label{section:vev}
As mentioned, at the central charge 24 of the Monster CFT, holomorphic factorization is possible, hence, the total central charge is 48, adding the central charge of an antiholomorphic counterpart. Therefore, the estimation for the VEV is a total energy of $2\times M_{\ell}\approx 250.8$ GeV. This is a naive estimation close to the 246.22 GeV VEV in the Standard Model. As discussed in the last point above, developing a nonzero VEV reduces the total rank by one unit, leading to a reduced total central charge of 47. Inserting the reduced central charge 47 instead of 48 (or 23.5 instead of 24) produces an improved estimation
\begin{equation}
\label{eq:10}
    VEV \approx \frac{47}{48} \times 250.8 \approx 245.7 \: GeV
\end{equation}
Our second observation is that this value is considerably closer to the Higgs VEV of $246.2$ GeV. This observation can also be expressed in the context of the Higgs self-coupling. In the Standard Model, the Higgs self-coupling constant is $\lambda=\frac{1}{2} {(\frac{M_H}{VEV})}^2\approx 0.129$. A naive estimation of $\frac{M_H}{VEV}$ is therefore the ratio of the central charge of the theory describing the Higgs mass to that of describing the corresponding VEV. Thus, based on the observations, it becomes the ratio of the central charge of the theory describing the Monster CFT to the total central charge. Therefore, a naive estimation of $\lambda$ is
\begin{equation}
\label{eq:11}
    \lambda =\frac{1}{2} {(\frac{c_{\:\mathbb M}}{c_T})}^2= \frac{1}{2} {(\frac{24}{48})}^2= \frac{1}{8} \approx 0.125
\end{equation}
Applying the rank reduction improves it to
\begin{equation}
\label{eq:12}
    \lambda = \frac{1}{2} {(\frac{24}{47})}^2 \approx 0.13
\end{equation}

\subsection{Gauge boson}
\label{section:gauage}
In our conjecture, based on the string theory, we assumed that 32 is the right dimension for the theory of quantum gravity. In this part, we observe that 32 precisely emerges as an output by analyzing $M_{\ell}$ at the Planck scale. Then we examine the consequence of this observation.
\subsubsection{Rank}
The Monster CFT does not have any spin-1 operators and Kac-Moody symmetry. Therefore, based on the CFT/(A)dS correspondences \cite{strominger2001ds} \cite{maldacena1999large}, it only describes pure gravity which does not include the gauge fields. A candidate theory that describes the gauge fields (that have more DOF) should possess a gauge group and higher dimensions than SU(2). The 3-dimensional SU(2) is the universal covering group of SO(3) that describes rotations in Euclidean 3-space. Let us promote the Euclidean 3-space to a Euclidean $d$-space and replace SO(3) with SO($d$). To find such a candidate, we calculate an unknown $d$ at which the boson mass of the same $\mathbb M$-symmetric vacuum configuration becomes exactly the reduced Planck mass, the critical scale of quantum gravity. That is, a dimension where $M_{\ell}=M_P$, which is equivalent to $m_\ell=1$. In this way, in Eq. \ref{eq:5}, the power of $\ell$ of the leading term in the order is promoted to $d$, and $\ell$ can be estimated solving $\ell^{d}\approx |\mathbb M|$. Hence, $\ell \approx |\mathbb M|^{(1/d)}$. Inserting this in Eq. \ref{eq:6},
\begin{equation}
\label{eq:13}
    m_\ell=\frac{48}{|\mathbb M|^{(1/d)}}
\end{equation}
Consequently, $m_\ell=1$ leads to $48^d\approx 8.08\times 10^{53}$. Solving for $d$ yields $d=32$ (at $d=32$, $m_s\approx 0.99$). 
\subsubsection{Massive gauge boson}
\label{section:massive gauge}
Based on the previous observation, the candidate gauge theory has a central charge 16. Therefore, the only lattices allowed by string theory are even unimodular root lattices of SO(32) and $E_8\times E_8$. Based on the Standard Model, massive gauge bosons gain mass via interaction with the Higgs field. Because the Higgs field (its microscopic DOF) is fixed, in interaction with it, particles gain mass proportional to their size/dimension '\textit{relative}' to the Higgs size/dimension. Thus, if the central charge 24 produces a $125.4$ GeV boson, what mass does a particle corresponding to the central charge $16$ gain? The answer is $\frac{16}{24}\times 125.4 \:GeV\approx 83.6$ GeV, where the ratio is the \textit{relative} size of 16 to 24. This answer (which can be obtained by replacing 24 by 16 in Eq. \ref{eq:6}) is a naive estimation. This estimation improves when the rank reduction is considered. The total central charge of 32, when reduced by one unit, becomes 31, giving rise to the reduced central charge of 15.5 . Accordingly, the improved estimation reads 
\begin{equation}
\label{eq:14}
    M_{G}={relative\:rank}\times M_H=\frac{15.5}{24}\times 125.4 \:GeV\approx 80.9\:GeV
\end{equation}
This is our third observation, where the estimation of the massive gauge boson is close to the measured W mass of $80.36$ GeV. This observation can also be expressed in the context of the Weak coupling constant. In the Standard Model, the weak coupling constant is $g=\frac{M_W}{VEV/2}\approx 0.641$. The naive estimation of the Higgs mass is proposed to be VEV/2. Thus, the naive estimation of $g$ is the ratio of the central charge of the theory describing the (massive) gauge boson to that of describing the Higgs boson, i.e., the 'relative rank' above. Hence, it becomes the ratio of the central charge of SO(32)-$E_8\times E_8$ theories to that of the Monster CFT
\begin{equation}
\label{eq:15}
    g= \frac{c_{\:SO(32)}}{c_{\:\mathbb M}}=\frac{16}{24}\approx 0.666
\end{equation}
Applying the rank reduction improves it to
\begin{equation}
\label{eq:16}
    g= \frac{15.5}{24}\approx 0.645
\end{equation}
Ultimately, let us examine the spin of this massive boson. The representation that describes massive spin-1 particles is the \textbf{3} representation of SU(2) which is the standard representation of SO(3). Particularly, a spin-$j$ boson has $2j+1$ DOF, hence, for $j=1$ and $j=0$, we have respectively 3 and 1 dimensions. Thus, the requirement that both fields share the same vacuum mass leads to the equality of length scales $|$Aut(spin-1)$| \approx |$Aut(spin-0)$|^{1/3}\approx \ell$. In other words, the order corresponding to the core symmetry of the spin-1 boson must be the cube root of that of the spin-0 boson. The quantum/core symmetry of SO(32) and $E_8\times E_8$ can be obtained by defining them over the finite field of length $q$ and then taking the limit $\lim_{q\to 1}$ of SO$_{32}$($q$) and in $E_8(q)\times E_8(q)$. It is known that this limit gives the Weyl group of the corresponding Lie group which is the automorphism group of their root lattice, i.e., $ \lim_{q\to 1}$ SO$_{32}$($q$)=W(SO(32)) and similarly $\lim_{q\to 1}$ $E_8(q)\times E_8(q)$=W($E_8\times E_8$) \footnote{Tits enlarges the analogy between $S_n$ and GL$_n(\mathbb F_q)$ to an analogy between the $\mathbb F_q$-rational points of a Chevalley group scheme $G$ and its Weyl group $W$ such that $ W = G(\mathbb F_1)$ \cite{tits1956analogues}. That is, $\mathbb F_1$ unveils the Weyl group which is the core or essence of the main group. He indicates that, in the $q=1$ limit, the finite geometry attached to $G(\mathbb F_q)$ will be the geometry of the Coxeter group $W$.} (for more details about this limit and its W($E_8\times E_8$) caluclation see \cite{khaki2024original}). Comparing the $|W(SO(32))|\approx 6.85\times 10^{17}$ and $|W(E_8\times E_8)|\approx4.85\times 10^{17}$ with $|\mathbb M|^{1/3}\approx9.31\times 10^{17}$, accords with the abovementioned requirement and confirms that such a massive boson should be spin-1.
\section*{Acknowledgements}
I appreciate the encouraging discussion with Masud Naseri.
\bibliographystyle{unsrt}
\bibliography{main.bib}

\begin{thebibliography}{10}

\bibitem{koren2020hierarchy}
Seth Koren.
\newblock The hierarchy problem: from the fundamentals to the frontiers.
\newblock {\em arXiv preprint arXiv:2009.11870}, 2020.

\bibitem{strominger2001ds}
Andrew Strominger.
\newblock The ds/cft correspondence.
\newblock {\em Journal of High Energy Physics}, 2001(10):034, 2001.

\bibitem{maldacena1999large}
Juan Maldacena.
\newblock The large-n limit of superconformal field theories and supergravity.
\newblock {\em International journal of theoretical physics}, 38(4):1113--1133, 1999.

\bibitem{frenkel1984natural}
Igor~B Frenkel, James Lepowsky, and Arne Meurman.
\newblock A natural representation of the fischer-griess monster with the modular function ${J}$ as character.
\newblock {\em Proceedings of the National Academy of Sciences}, 81(10):3256--3260, 1984.

\bibitem{conway1979monstrous}
John~H Conway and Simon~P Norton.
\newblock Monstrous moonshine.
\newblock {\em Bulletin of the London Mathematical Society}, 11(3):308--339, 1979.

\bibitem{gannon2007moonshine}
Terry Gannon.
\newblock {\em Moonshine beyond the Monster: The bridge connecting algebra, modular forms and physics}.
\newblock Cambridge University Press, 2007.

\bibitem{harrison2022snowmass}
Sarah~M Harrison, Jeffrey~A Harvey, and Natalie~M Paquette.
\newblock Snowmass white paper: Moonshine.
\newblock {\em arXiv preprint arXiv:2201.13321}, 2022.

\bibitem{sirunyan2020measurement}
Albert~M Sirunyan et~al.
\newblock A measurement of the higgs boson mass in the diphoton decay channel.
\newblock {\em Physics Letters B}, 805:135425, 2020.

\bibitem{aad2023measurement}
Georges Aad et~al.
\newblock Measurement of the higgs boson mass with h→ $\gamma$$\gamma$ decays in 140 fb- 1 of s= 13 tev pp collisions with the atlas detector.
\newblock {\em Physics Letters B}, 847:138315, 2023.

\bibitem{witten2001quantum}
Edward Witten.
\newblock Quantum gravity in de sitter space.
\newblock {\em arXiv preprint hep-th/0106109}, 2001.

\bibitem{witten2007three}
Edward Witten.
\newblock Three-dimensional gravity revisited.
\newblock {\em arXiv preprint arXiv:0706.3359}, 2007.

\bibitem{wilson1999atlas}
Robert~A Wilson et~al.
\newblock Atlas of finite group representations, 1999.

\bibitem{buttazzo2013investigating}
Dario Buttazzo et~al.
\newblock Investigating the near-criticality of the higgs boson.
\newblock {\em Journal of High Energy Physics}, 2013(12):1--49, 2013.

\bibitem{eguchi2011notes}
Tohru Eguchi, Hirosi Ooguri, and Yuji Tachikawa.
\newblock Notes on the $k3$ surface and the mathieu group m 24.
\newblock {\em Experimental Mathematics}, 20(1):91--96, 2011.

\bibitem{taormina2015twist}
Anne Taormina and Katrin Wendland.
\newblock A twist in the ${M}_ {24}$ moonshine story.
\newblock {\em Confluentes Mathematici}, 7(1):83--113, 2015.

\bibitem{tits1956analogues}
Jacques Tits.
\newblock Sur les analogues alg{\'e}briques des groupes semi-simples complexes.
\newblock In {\em Colloque d’algebre sup{\'e}rieure, tenua Bruxelles du}, volume~19, pages 261--289, 1956.

\bibitem{khaki2024original}
Seyed Khaki.
\newblock Original {$\FF_{1}$} in emergent spacetime.
\newblock {\em arXiv preprint arXiv:2401.07822}, 2024.

\bibitem{cohn2017sphere}
Henry Cohn, Abhinav Kumar, Stephen Miller, Danylo Radchenko, and Maryna Viazovska.
\newblock The sphere packing problem in dimension 24.
\newblock {\em Annals of mathematics}, 185(3):1017--1033, 2017.

\bibitem{cohn2022universal}
Henry Cohn et~al.
\newblock Universal optimality of the ${E}_8$ and leech lattices and interpolation formulas.
\newblock {\em Annals of Mathematics}, 196(3):983--1082, 2022.

\bibitem{yang1952statistical}
Chen-Ning Yang and Tsung-Dao Lee.
\newblock Statistical theory of equations of state and phase transitions. i. theory of condensation.
\newblock {\em Physical Review}, 87(3):404, 1952.

\bibitem{lee1952statistical}
Tsung-Dao Lee and Chen-Ning Yang.
\newblock Statistical theory of equations of state and phase transitions. ii. lattice gas and ising model.
\newblock {\em Physical Review}, 87(3):410, 1952.

\bibitem{suyama2002properties}
Takao Suyama.
\newblock Properties of string theory on kaluza-klein melvin background.
\newblock {\em Journal of High Energy Physics}, 2002(07):015, 2002.

\bibitem{hohn2010group}
Gerald H{\"o}hn.
\newblock The group of symmetries of the shorter moonshine module.
\newblock In {\em Abh. Math. Semin. Univ. Hambg.}, volume~80, pages 275--283. Springer, 2010.

\bibitem{conway2013sphere}
John~Horton Conway and Neil James~Alexander Sloane.
\newblock {\em Sphere packings, lattices and groups}, volume 290.
\newblock Springer Science \& Business Media, 2013.

\bibitem{conway1988low}
John~Horton Conway and Neil~JA Sloane.
\newblock Low-dimensional lattices. ii. subgroups of ${GL (n, \mathbb Z}$).
\newblock {\em Proceedings of the Royal Society of London. A. Mathematical and Physical Sciences}, 419(1856):29--68, 1988.

\end{thebibliography}
\end{document}